\documentclass[pre,twocolumn,amssymb,amsmath,showpacs]{revtex4}

\usepackage{graphicx}
\usepackage{color}
\usepackage{epsfig}
\usepackage{latexsym}
\usepackage{bm}
\usepackage{ulem}

\usepackage{color}
\definecolor{dgreen}{rgb}{0,0.7,0}

\begin{document}

\renewcommand{\ni}{{\noindent}}
\newcommand{\dprime}{{\prime\prime}}
\newcommand{\be}{\begin{equation}}
\newcommand{\ee}{\end{equation}}
\newcommand{\bea}{\begin{eqnarray}}
\newcommand{\eea}{\end{eqnarray}}
\newcommand{\nn}{\nonumber}
\newcommand{\bk}{{\bf k}}
\newcommand{\bQ}{{\bf Q}}
\newcommand{\q}{{\bf q}}
\newcommand{\s}{{\bf s}}
\newcommand{\bN}{{\bf \nabla}}
\newcommand{\bA}{{\bf A}}
\newcommand{\bE}{{\bf E}}
\newcommand{\bj}{{\bf j}}
\newcommand{\bJ}{{\bf J}}
\newcommand{\bs}{{\bf v}_s}
\newcommand{\bn}{{\bf v}_n}
\newcommand{\bv}{{\bf v}}
\newcommand{\la}{\langle}
\newcommand{\ra}{\rangle}
\newcommand{\dg}{\dagger}
\newcommand{\br}{{\bf{r}}}
\newcommand{\brp}{{\bf{r}^\prime}}
\newcommand{\bq}{{\bf{q}}}
\newcommand{\hx}{\hat{\bf x}}
\newcommand{\hy}{\hat{\bf y}}
\newcommand{\bS}{{\bf S}}
\newcommand{\cU}{{\cal U}}
\newcommand{\cD}{{\cal D}}
\newcommand{\bR}{{\bf R}}
\newcommand{\pll}{\parallel}
\newcommand{\sumr}{\sum_{\vr}}
\newcommand{\cP}{{\cal P}}
\newcommand{\cQ}{{\cal Q}}
\newcommand{\cS}{{\cal S}}
\newcommand{\ua}{\uparrow}
\newcommand{\da}{\downarrow}
\newcommand{\red}{\textcolor {red}}
\newcommand{\blu}{\textcolor {blue}}
\newcommand{\1}{{\oldstylenums{1}}}
\newcommand{\2}{{\oldstylenums{2}}}
\newcommand{\mDelta}{\varepsilon}
\newcommand{\m}{\tilde m}
\def\lsim {\protect \raisebox{-0.75ex}[-1.5ex]{$\;\stackrel{<}{\sim}\;$}}
\def\gsim {\protect \raisebox{-0.75ex}[-1.5ex]{$\;\stackrel{>}{\sim}\;$}}
\def\lsimeq {\protect \raisebox{-0.75ex}[-1.5ex]{$\;\stackrel{<}{\simeq}\;$}}
\def\gsimeq {\protect \raisebox{-0.75ex}[-1.5ex]{$\;\stackrel{>}{\simeq}\;$}}

\title{Einstein relation and hydrodynamics of nonequilibrium  mass
transport processes}

\author{Arghya Das$^{1}$, Anupam Kundu$^{2}$ and Punyabrata Pradhan$^{1}$}

\affiliation{$^1$Department of Theoretical Sciences, S. N. Bose
National Centre for Basic Sciences, Block-JD, Sector-III, Salt
Lake, Kolkata 700106, India \\ $^2$International Centre for
Theoretical Sciences, TIFR, Bangalore 560012, India}

\begin{abstract}

\noindent{We derive hydrodynamics of paradigmatic conserved-mass transport processes on a ring. The systems, governed by chipping, diffusion and coalescence of masses,  eventually reach a nonequilibrium steady state, having nontrivial correlations, with steady-state measures in most cases not known. In these
processes, we analytically calculate two transport coefficients,
bulk-diffusion coefficient and conductivity. Remarkably, the two transport coefficients obey an equilibriumlike Einstein relation even when the microscopic dynamics violates detailed balance and systems are far from equilibrium. Moreover, we show, using a
macroscopic fluctuation theory, that the probability of
large deviation in density, obtained from the above hydrodynamics, is
in complete agreement with the same derived earlier in [Phys. Rev.
E {\bf 93}, 062135 (2016)] using an additivity property. }

\typeout{polish abstract}

\end{abstract}


\maketitle

\section{Introduction}

\noindent

The Einstein relation (ER) \cite{Einstein}, also known as the
Einstein-Smoluchowski relation, is a celebrated equality in
equilibrium physics. It connects, quite unexpectedly, two
seemingly unrelated transport coefficients, bulk-diffusion
coefficient $D(\rho)$ and conductivity $\chi(\rho)$, as $D(\rho) =
\chi(\rho)/{\sigma^2_{eq}(\rho)}$ where $\sigma^2_{eq}(\rho) =
\lim_{v \rightarrow \infty}(\langle n_v^2 \rangle_{eq} - \langle
n_v \rangle^2_{eq})/v$ is scaled variance of particle-number $n_v$
in a sub-volume $v$ still much smaller than the system volume and $\rho$ is local number density; angular bracket $\langle \cdot \rangle_{eq}$ denotes equilibrium average.   
Here the diffusion coefficient $D(\rho)$ is defined from Fourier's law for diffusive current $J_D = -D(\rho) \partial
\rho/\partial x$ where $\partial \rho/\partial x$ is spatial
density gradient in a particular direction, say along $x-$axis.
The conductivity $\chi(\rho)$ is defined from Ohm's law for
drift current $J_d = \chi(\rho) F/k_BT$, due to a small external biasing force
$F$ also along $x-$axis, with $k_B$ and $T$ being the Boltzmann
constant and temperature, respectively.

For systems in equilibrium where detailed balance is obeyed, the
ER is universal, irrespective of the details of
inter-particle interactions or that of whether the systems are
liquids or gases, etc. Indeed, the ER is one of the
earliest known forms of a more general class of equilibrium
fluctuation relations, collectively called fluctuation-dissipation
theorems (FDTs); the FDTs can be proved using linear-response
theory around equilibrium state having the Boltzmann-Gibbs
distribution \cite{Green-Kubo}.

However, systems having a nonequilibrium  steady state (NESS),
which is arguably the closest counterpart to equilibrium,
generally do not have such relations. Because, unlike in
equilibrium, they violate detailed balance and usually cannot be
described by the Boltzmann-Gibbs distribution. In fact, in most
cases, microscopic probability weights in the steady state are
{\it not} known. Quite interestingly recent studies
\cite{Jona-Lasinio_PRL1996, Lebowitz_Spohn_JSP1999,
Bertini_PRL2001, Nagel, Sasa_PRE2003, Eyink, Bechinger_PRL2007, 
Bertin_PRL2006, Pradhan_PRL2010, Chatterjee_PRL2014, Das_PRE2015, 
Das_PRE2016, Bertini_review} 
have indicated that, even in NESSs, there can be fluctuation 
relations analogous to the FDTs in equilibrium. In particular, the 
ER has been found, mostly numerically, in several model systems 
\cite{Sasa1, Sasa2, Hurtado_PRE2012} having a NESS.

The ER involves two bulk transport coefficients $D(\rho)$ and $\chi(\rho)$, defined on a macroscopic level from the two phenomenological laws of transport - Fourier's and Ohm's law.
One way to understand such macroscopic phenomenological fluctuation relations 
is to derive, from microscopic dynamics, a hydrodynamic description 
of the systems on a large space and time scales. 
However, such a task, for classical deterministic (or quantum) 
dynamics, is quite difficult. On the other hand, for systems 
governed by stochastic dynamics, the problem of deriving 
hydrodynamics is comparatively easier and, recently, there has been 
considerable progress made in this direction \cite{ELS_CMP1990, 
Book_Spohn, Book_Kipnis_Landim}. 
However, for 
stochastic systems having a NESS, the steady-state probability 
weights are not always known and tackling the problem analytically in such systems, especially when there are nonzero finite spatial correlations,
remains to be a challenging one \cite{Nakamura_Sasa}. Perhaps not 
surprisingly, so far there are not many nonequilibrium interacting-particle systems for 
which exact hydrodynamic descriptions, presumably the first step towards exploring fluctuation relations such as the ER, have been derived. In fact, the difficulty arises primarily because fluctuation, diffusion coefficient and conductivity, which would appear in ER (if any) in such systems, must be calculated in a steady state far from equilibrium, not in or around an equilibrium state.

Here, we study a broad class of nonequilibrium conserved-mass  transport
processes on a ring. These processes are governed by chipping,
diffusion and coalescence of neighboring masses, with total mass in 
the system being conserved, and have become paradigm in 
nonequilibrium statistical physics of driven many-particle systems 
\cite{Aldous, Ferrari}. Indeed, throughout the last couple of 
decades, they have
been explored intensively to model a huge variety of natural
phenomena, such as, formation of clouds \cite{cloud} and gels
\cite{gel, Rajesh_JSP2000}, force fluctuation in packs of granular beads 
\cite{Majumdar_Science1995, Majumdar_PRE1996}, transport of 
energy in solids \cite{KMP}, dynamics of interacting particles on
a ring \cite{Krug_JSP2000}, self-assembly of molecules in organic 
and inorganic materials \cite{lipid, Vledouts_RSPA2016}, and distribution of wealth in a society
\cite{wealth_review}, etc.

In this paper, we derive hydrodynamics of the  above mentioned
one-dimensional conserved-mass transport processes, which
have nontrivial spatial correlations (nonzero and finite), with their steady-state
weights in most cases not known. For these processes, we
explicitly calculate the two transport coefficients
as a function of local mass density $\rho$ $-$ the bulk-diffusion
coefficient $D(\rho)$ and the conductivity $\chi(\rho)$, which
characterize the hydrodynamics. Remarkably, we found that, 
for this class of models, the two transport coefficients satisfy an 
equilibriumlike Einstein relation, 
\be D(\rho) = \frac{\chi(\rho)}{\sigma^2(\rho)},
\label{ER} 
\ee 
where 
\be \sigma^2(\rho) = \lim_{v \rightarrow
\infty} \frac{\langle m^2 \rangle - \langle m \rangle^2}{v},
\label{sigma} 
\ee 
is scaled variance of mass $m$ in a large subsystem (much smaller than the system) of volume $v$ with $\rho = \langle m \rangle/v$ is average local mass density.
The diffusion coefficient $D(\rho)$ and the conductivity $\chi(\rho)$ are suitably defined on a hydrodynamic level from diffusive current $J_D = -D(\rho) \partial \rho/\partial x$ and drift current $J_d = \chi(\rho) F$, respectively, where $\partial \rho/\partial x$ is gradient in local mass density and $F$ is the magnitude of a small biasing force  coupled locally to conserved mass variable and applied in a particular direction. 
For all the processes considered in this paper, we find bulk
diffusion coefficient $D(\rho) = \rm{const.}$ and conductivity
$\chi(\rho) \propto \rho^2$, indicating that the processes, on
hydrodynamic level, belong to the class of Kipnis-Marchioro-Presutti
(KMP) processes on a ring \cite{KMP}. Moreover, we use the two
transport coefficients to find probabilities of large
deviations of mass in a subsystem in the framework of recently
developed macroscopic fluctuation theory (MFT)
\cite{Bertini_PRL2001, Bertini_review}. The mass large-deviation
functions (LDFs) completely agree with that in Refs.
\cite{Chatterjee_PRL2014, Das_PRE2016}, which were derived earlier
using an additivity property.

The paper is organised as follows. In Sec. \ref{GenCon}, we
discuss general aspects of conserved-mass transport processes. In
Sec. \ref{Theory_Results}, we present a linear-response analysis around a nonequilibrium steady state, which is implemented to calculate the transport coefficients in the model-systems discussed later. We introduce, in Sec. \ref{SMCM} (symmetric versions) and Sec. \ref{AMCM} (asymmetric versions), a broad class of conserved-mass transport processes (called models I, II and III) and derive hydrodynamics of these systems in terms of two transport coefficients - the diffusion coefficient and the 
conductivity. In Sec. \ref{Sec_LDF} and \ref{AMCM}, we
discuss how the density large deviation functions in all these models can be calculated
using a macroscopic fluctuation theory. In Sec. \ref{Summary}, we summarize with some concluding remarks.

\section{General considerations and motivations}
\label{GenCon}

Let us first discuss some general  aspects of fluctuations in
steady states and their connection to hydrodynamics in the context
of recently obtained results in conserved-mass transport processes
\cite{Chatterjee_PRL2014, Das_PRE2015, Das_PRE2016}.  The
conserved-mass transport processes are defined on a one
dimensional periodic lattice of $L$ sites, with a continuous mass
variable $m_i \ge 0$ at site $i \in \{1, 2, \dots L\}$
\cite{Krug_JSP2000, Rajesh_JSP2000, Zielen_JSP2002, Mohanty_JSTAT2012}. They are
governed by dynamical rules, such as, chipping or fragmentation,
diffusion and coalescence of neighboring masses, which eventually
lead to a nonequilibrium steady state. Under these dynamical
rules, total mass $M=\sum_{i=1}^L m_i$ in the system remains
conserved. Though these processes are governed by simple dynamical
rules, they usually have nontrivial spatial correlations in the
steady states. That is why, even in one dimension (which is the
case considered here), the {\it exact} steady-state probability
weights for the microscopic configurations, except for a few
special cases \cite{Majumdar_PRE1996, Krug_JSP2000,
Rajesh_JSP2000, Zielen_JSP2002}, are not yet known.

In this paper, we study several generalized versions of the above mentioned mass transport
processes \cite{Das_PRE2016}, which we call Model I, Model II
and Model III. In the symmetric versions of the models (see Sec. \ref{SMCM}), mass transfers take place, without any preference, to the right or (and) to the left
nearest neighbor(s); consequently, net mass currents are zero in
the nonequilibrium steady states. However, as shown later, the systems with the symmetric transfers still remain far from equilibrium 
as the dynamics in the configuration space violates Kolmogorov criterion and thus also detailed balance \cite{kolmogorov}. For asymmetric mass transfers (see Sec. \ref{AMCM}), the violation is quite evident as there would be nonzero mass current in the systems. Kolmogorov criterion, which provides a necessary and sufficient condition for detailed balance to hold in a system, says the following. If, for each and every possible loop generated by the dynamics in the configuration space, the probability of a forward path and that of the corresponding reverse path are equal, detailed balance is satisfied, and {\it vice versa}.  As a consequence, if a reverse path corresponding to a forward path in a particular transition in the configuration space does not exist, it suffices to say that Kolmogorov criterion, and therefore detailed balance, is violated.
Indeed, in the absence of the knowledge of exact steady-state measures in these mass transport processes, Kolmogorov criterion helps one to check whether detailed balance is satisfied or not.

At a coarse-grained level where  one divides such a system of
volume $V$ into $\nu=V/v$ subsystems, each of volume $v \ll V$,
one could however have a simpler description. Provided that the
subsystem sizes are large compared to the microscopic spatial
correlation length but much smaller than the size of the full
system, one expects that the system would possess an additivity
property \cite{Eyink, Bertin_PRL2006, Pradhan_PRL2010}, which states that large
subsystems are statistically almost independent. That is, the {\it
steady-state} joint subsystem mass distribution ${\cal P}[\{ M_1,
M_2, \dots, M_\nu \}]$, with $M_k$ being mass in $k$th subsystem,
can be approximately written in a product form, except for a
constraint of global mass conservation. In other words, the joint
subsystem mass distribution can be expressed in terms of subsystem
weight factor $W_v(M_k)$, \bea {\cal P}[\{M_k\}] \simeq
\frac{\prod_k W_v(M_k)}{Z(V, M)} \delta \left( \sum_k M_k - M
\right), \label{additivity} \eea where $Z$ is the normalization
constant. For large subsystem size, the weight factor $W_v(M_k)$
can be characterized by a large deviation `density' function
$f(\rho_k)$ (or `rate' function; also sometimes called
`nonequilibrium free energy' density) as $W_v(M_k) \simeq
\exp[-vf(\rho_k)]$ where $\rho_k = M_k/v$ is fluctuating subsystem
mass density \cite{Das_PRE2015}. The immediate consequence of
additivity is that the function $f(\rho)$ is related to the scaled
variance $\sigma^2(\rho)$ [as defined in eq. (\ref{sigma})]
of subsystem mass through a fluctuation-response relation (FR)
\cite{Eyink, Bertin_PRL2006, Pradhan_PRL2010, Chatterjee_PRL2014, 
Das_PRE2015, Das_PRE2016}, analogous to equilibrium
fluctuation-dissipation theorems, \bea f''(\rho) = \frac{d \mu}{d
\rho} = \frac{1}{\sigma^2(\rho)}, \label{f-sigma-rela} \eea where
$\mu(\rho) = f'(\rho)$ is defined to be a chemical potential and
$\rho = \langle \rho_k \rangle$ is local mass density. Now,
instead of subsystem mass variables $\{M_k\}$, additivity property
[eq. (\ref{additivity})] can be written in terms of subsystem
density variables $\{\rho_k = M_k/v \}$, or equivalently, in terms
of coarse-grained fluctuating density profile $\{\rho(x)\}$ 
in the system. Then, one can write the joint subsystem density
distribution, or large-deviation probability of a given density
profile $\{\rho(x)\}$, as $${\cal P}[\rho(x)] \simeq
e^{-\mathcal{F} [\rho(x)]},$$ where $\mathcal{F} [\rho(x)]$ is
called large deviation function (LDF). In the mass-transport
processes considered here, as the functional form of the scaled
variance $\sigma^2(\rho) = \rho^2/\eta$ with $\eta$ being a model
dependent parameter [e.g., see eq. (\ref{sigma-model1})], the LDFs
can be calculated by using additivity and the FR (Eqs.
\ref{additivity} and \ref{f-sigma-rela}) \cite{Chatterjee_PRL2014,
Das_PRE2016}. In fact, the LDFs have been previously shown to have
the following form, \be \mathcal{F}[\rho(x)] = \int_V dx~
\{f(\rho)-f(\rho_0) - \mu(\rho_0)(\rho-\rho_0) \},\label{LDF-a}
\ee where \bea f(\rho) = - \eta \ln \rho, \label{f-rho}
\\
\mu(\rho) = f'(\rho) = - \frac{\eta}{\rho},
\label{mu-rho}
\eea
with $\mu(\rho)$ an equilibriumlike chemical potential and $\rho_0 = M/V$ the global mass density \cite{Chatterjee_PRL2014, Das_PRE2016}. The FR in eq. (\ref{f-sigma-rela}) can be verified from eq. (\ref{LDF-a}). Moreover, in this case, the probability distribution function $P_v(m)$ of mass $m$ in a subsystem of volume $v$ can be obtained as
\be 
P_v(m) \propto m^{v \eta -1} e^{-\eta m/\rho},
\label{Pvm}
\ee
which is gamma distribution \cite{Chatterjee_PRL2014, Das_PRE2016}.

Thus, additivity property helps  one to construct a statistical
mechanical framework in these conserved-mass transport processes,
through a free energy density $f(\rho)$ and a chemical
potential $\mu(\rho)$, which however describes only the static
properties of {\it steady-state} mass fluctuations. At this
stage, one could ask whether the above LDFs can be derived in a
dynamical setting. To address this issue, here we formulate,
within recently developed macroscopic fluctuation theory (MFT)
\cite{Bertini_PRL2001, Bertini_review}, a statistical mechanical description of
fluctuations for these processes. The
formulation provides a dynamical description of mass fluctuations
at a coarse-grained level, i.e., a fluctuating hydrodynamics valid
in large length and time scales [see eq. (\ref{FHD})].

Since mass remains conserved locally under the microscopic
evolution, one must keep the mass conservation valid also at the
hydrodynamic scales. Therefore, the hydrodynamic equation
must be written in the form of a continuity equation, \be
\partial_\tau \rho (x, \tau) + \partial_x J(\rho(x,\tau)) = 0,
\label{continuity} \ee which  governs the time evolution of
density field $\rho(x,\tau)$ with $x$ and $\tau$ being suitably
rescaled position and time, respectively. Since, the class of
processes we consider here are of `gradient type' (i.e., local diffusive current can be expressed as a gradient in local observables)
\cite{Jona-Lasinio_PRL1996} with respect to their microscopic
evolutions, one would expect a nonlinear hydrodynamics in the
diffusive scaling limit, where the current $J(\rho(x,\tau))$ is
sum of two parts $J =J_D + J_d$. The first part $J_D = -D(\rho)
\partial_x \rho$ is the diffusive current with $D(\rho)$ being the
diffusion coefficient and the second part  $J_d(\rho,\tau) = \chi
(\rho) F$ is the drift current due to a small slowly varying
biasing field $F(x)$ (conjugate to conserved mass variable) with
$\chi(\rho)$ being the conductivity.

According to the hydrodynamic  equation (\ref{continuity}),
along with a constitutive relation for the current $J(\rho)=-D(\rho)
\partial_x \rho + \chi(\rho) F$, the density field $\rho(x, \tau)$
evolves deterministically in time. However, to study any dynamical
aspects of fluctuations, one requires to add a suitable noise
term. Clearly, as the noise in this case should maintain the local
mass conservation, one must add a noise term $\zeta$ to the deterministic part
of the current $J(x,\tau) \rightarrow J(x,\tau) + \zeta(x, \tau)$,
making the total current now a fluctuating one. But the question
here is what properties the noise $\zeta$ would have. As we see
later within MFT [see eq. (\ref{FHD})], the fluctuating part
$\zeta$ of the total current can be represented in terms of a weak
multiplicative Gaussian white noise, whose strength explicitly
depends on the conductivity $\chi(\rho)$. So the problem of
formulating a theory of mass fluctuations in these processes
essentially boils down to finding the functional dependence of the
diffusion coefficient $D(\rho)$ and the conductivity $\chi(\rho)$
on density $\rho$.

In the following section, we  explicitly calculate the two
transport coefficients, $D(\rho)$ and $\chi(\rho)$, in a broad
class of conserved-mass transport processes. Remarkably, in all
cases studied here, we find that the two transport coefficients
obey an Einstein relation eq. (\ref{ER}). We present below the
details of computations for different models separately.

\section{Theory: Linear Response around nonequilibrium steady states }
\label{Theory_Results}

Before proceeding to the calculations of the transport 
coefficients in the nonequilibrium mass transport processes
mentioned in the previous section, we first present a proof of the
Einstein relation (ER), which is valid in or, strictly speaking, around equilibrium state of a system, in the limit of an external force vanishingly small. In equilibrium, an
external force field $\vec{F}$ (here taken to be constant, for
simplicity), or equivalently an external potential, can be
directly related to chemical potential of the system. For example,
consider a one dimensional system whose two halves are kept at two
different external potentials, say, first half at potential $V_1$
and second half at potential $V_2$ where $V_2 - V_1 = \Delta V = -
\int F dx$ with the force field $\vec{F} = F\hat{x}$. The fact
that effective chemical potentials of the two halves equalize
implies $$\mu(\rho_1) + V_1 = \mu(\rho_2) + V_2,$$ where $\rho_1$
and $\rho_2$ are densities of the first and second halves,
respectively, $\mu(\rho) = df/d\rho$ is chemical potential
(canonical) and $f(\rho)$ free energy density (canonical) in the
absence of any external potential. In other words, across a
spatial interval $\Delta x$, we have the following relation
$\Delta \mu/\Delta x = - \Delta
V/\Delta x = F$, or \be \frac{d\mu}{dx} = F, \label{mu-F} \ee in
the limit of $\Delta x \rightarrow 0$. Now, in the limit of small
force $F \rightarrow 0$, drift current $J_d=\chi(\rho) F$ due to the force $F$ and
the diffusion current $J_D = - D(\rho) d\rho/dx$ must balance each
other so that there is no net current in the system. That is, we must
have $J_d + J_D = 0$, which, along with the equality $F = d\mu/dx =
(d\mu/d\rho) (d\rho/dx)$ [from eq. (\ref{mu-F})] and the equilibrium fluctuation-response relation between compressibility 
and fluctuation $d\rho/d\mu = \sigma^2(\rho)$ 
[eq. (\ref{f-sigma-rela})], immediately leads to the ER.

On the other hand, in nonequilibrium, though  detailed balance
is violated on a microscopic level, the macroscopic mass current 
in the steady state could still be zero, e.g., in the case of the 
mass-transport processes with symmetric mass transfer rules. In that case, one would perhaps expect, for a 
suitably chosen biasing force, an ER even in nonequilibrium. 
Interestingly, we see later that an ER holds in the cases of both symmetric and asymmetric mass transfers.
The issue essentially revolves around the crucial question whether eq. (\ref{mu-F}) would hold in such cases, which could be addressed 
by checking if there is an ER. In fact, provided
it holds, an ER would then imply a LDF of the form as in eq.
(\ref{LDF-a}) where $f''(\rho) = D(\rho)/\chi(\rho)$ (see Sec.
\ref{Sec_LDF} for a more rigorous discussion).

To explore the issue further, we perform a linear-response analysis of
the conserved-mass transport processes in the presence of a small
constant biasing force field $\vec{F}=F \hat{x}$, which is now applied
in the system, with $\hat{x}$ being a unit vector along $+ve$ $x$
axis. The force field $\vec{F}$, somewhat like a  gravitational one, is conjugate to the conserved-mass variables (external force is coupled to
local masses at the individual sites) and is chosen as following.
 The biasing force
$\vec{F}$ modifies the original mass transfer rates $c_{i \rightarrow
j}$, from site $i$ to $j$, to biased rates $c^F_{i \rightarrow j}$
(which are now effectively asymmetric) \cite{Bertini_review}, \be
c^F_{i \rightarrow j} = c_{i \rightarrow j} \Phi(\Delta e_i),
\label{bias1} \ee where $\Phi(\Delta e_i)>0$ is non-negative
function of \be \Delta e_i = \Delta m_{i \rightarrow j}
(\vec{F}.\delta \vec{x}_{ij}). \ee The quantity $\Delta e$ can be
physically interpreted as extra energy cost (due to the biasing force $\vec{F}$), for transferring or
displacing mass $\Delta m_{i \rightarrow j}$ from site $i$ to $j$
in a particular direction with the mass displacement vector
$\delta \vec{x}_{ij} = (j-i) a \hat{x}$ and $a$ being the lattice
constant. We explicitly write the lattice constant, which would be
required later for taking diffusive scaling limit. Clearly,
$\Phi|_{F=0} = 1$ as $c^{F=0}_{i \rightarrow j} = c_{i \rightarrow
j}.$

In the case of only nearest-neighbor mass transfer  (more
generalized version is described below), the mass displacement
vector $\delta \vec{x}_{i j}$ can take, depending on the direction of 
the mass transfer, one of the two values $\delta \vec{x} = \pm a 
\hat{x}$. Consequently, the form of rates in eq. (\ref{bias1}) makes the modified forward 
and backward mass-transfer rates across a bond asymmetric and 
therefore induces a small net current in the system.

To check the ER, we consider, somewhat analogous  to equilibrium,
the function $\Phi$ to have a form $\Phi(\Delta e) = \exp(\Delta
e/2)$ \cite{Bertini_review}. However, note that, in the following
linear analysis for small force $F$  where we require only the
leading order term ${\cal O}(F)$ (or ${\cal O}(\Delta e)$), the
whole analysis goes through even for a general functional form of
$\Phi$. We expand $\Phi$ in ${\cal O}(F)$, \be \Phi (\Delta e_i)
\simeq 1 + \left[ \frac{d \Phi}{d (\Delta e)} \right]_{\Delta e=0}
\Delta e_i = 1 + \frac{1}{2} \Delta m_{i \rightarrow j}
(\vec{F}.\delta \vec{x}_{ij}). \label{phi-expn} \ee 
For example, see the biased mass-transfer rates
$c^F_{i \rightarrow j}$ as in eqs. (\ref{rate1-model-II-RSU}) and
(\ref{rate2-model-II-RSU}). In the above equation, without
any loss of generality, we put $2[{d \Phi}/{d \Delta e}]_{\Delta
e=0}=1$, which essentially implies a rescaling of the applied force 
$F \rightarrow [2{d \Phi}/{d (\Delta e)}]_{\Delta e=0} \times F$.

It is possible that several fractions  $\Delta m_{i_n \rightarrow
j_{n'}}$, where $n=1, 2, \dots, K$ and $n'=1, 2, \dots, K'$, of
masses from $K$ number of sites $\{i_n\} \equiv i_1, i_2, \dots
i_K$ are transferred, at the same instant of time, to $K'$ number
of sites $\{j_n'\} \equiv \{j_1, j_2, \dots j_{K'} \}$.
For example, see the modified rates for Model I in eq.
(\ref{rate-model-I-RSU})
where $K=1$ and $K'=2$ and in eq. (\ref{bias3})
where $K=K'=L$. The original rate $c_{\{i_n\} \rightarrow
\{j_{n'}\} }$, for mass transfer from sites $\{i_n\}$ to
$\{j_{n'}\}$, and the corresponding modified biased rate
$c^F_{\{i_n\} \rightarrow \{j_{n'}\} }$ are related as \be
c^F_{\{i_n\} \rightarrow \{j_{n'}\} } = c_{\{i_n\} \rightarrow
\{j_{n'}\} } \Phi(\Delta e), \label{bias2} \ee where the total
extra energy cost, due to the biasing, can be written by summing
over all individual energy costs corresponding to each and every
pair $\langle n,n' \rangle$ of departure site $n$ and destination site $n'$ as
\be \Delta e = \sum_{\langle n,n'\rangle} \Delta m_{i_n
\rightarrow j_{n'}} (\vec{F}.\delta \vec{x}_{i_n j_{n'}}).
\label{De} \ee In the next, we use this modified biased rate
$c^F_{\{i_n\} \rightarrow \{j_{n'}\} } $ [as in eq. (\ref{bias2})]
alongwith eqs. (\ref{phi-expn}) and (\ref{De}) for the three
models (I, II, III) to derive a hydrodynamic equation like in Eq.~
\ref{continuity} and hence, in turn, we compute the diffusivity
$D(\rho)$ and the conductivity $\chi(\rho)$.

\section{models and results: symmetric mass transfers}
\label{SMCM}

In this section, we define the symmetric versions of the models, first in the absence of any biasing force, where masses are transferred symmetrically, without any preferential direction, to the nearest neighbors. Consequently, there is no net mass current in the systems. However, it is important to note that, even in that case, detailed balance condition is still not satisfied. In fact, it would be quite instructive to explicitly show that, for generic values of parameters in the models, Kolmogorov criterion and therefore detailed balance is strongly violated, in the sense that, for a transition (say, forward) from one  configuration to another while mass being transferred from a site to its neighbor, the corresponding reverse path of transition may not exist. 

Therefore, even in the absence of any biasing force, the system eventually reaches a steady state, which is inherently far from equilibrium, and cannot be described by the equilibrium Boltzmann-Gibbs distribution. To calculate conductivity in such a nonequilibrium steady state, we need to apply a biasing (constant, for simplicity) force field, which would essentially modify the original mass-transfer rates in the systems, inducing a mass current, and then we calculate the current in the limit of biasing force being small.

\subsection{Model I}
\label{Model-I}

This particular class of models has been introduced to  study mass transport
processes accounting for stickiness of masses while fragmenting
and diffusing \cite{Mohanty_JSTAT2012}. These processes are
variants of various previously studied mass transport processes, such
as random average processes (RAP), etc. \cite{Ferrari, Krug_JSP2000, 
Rajesh_JSP2000}.

\subsubsection{Random Sequential Update}

In Model I with random sequential update (RSU),  three sites are
updated simultaneously where two random fractions of the
chipped-off mass from site $i$ are shared randomly with the
nearest neighbour sites $i-1$ and $i+1$.  The stochastic time
evolution of mass $m_i(t)$ at time $t$ after an infinitesimal time
$dt$ can be written as \be m_i(t+dt) = \label{model-1}
\begin{cases}
\lambda m_i(t) & \mbox{prob.} ~dt \\
m_i(t)+ \tilde{\lambda}r_{i-1}m_{i-1}(t)  & \mbox{prob.} ~dt\\
m_i(t)+\tilde{\lambda}\tilde{r}_{i+1}m_{i+1}(t) & \mbox{prob.} ~dt\\
m_i(t) & \mbox{prob.}~ (1-3dt)
\end{cases}
\ee  where $r_j\in ({0,1})$s are  independent and  identically
distributed (i.i.d.) random variables, having a probability
density  $\phi(r)$ and $\tilde{\lambda} = 1 - \lambda$ and
$\tilde{r}_{i+1}=1-r_{i+1}$. Throughout the paper, we denote the
first and the second moments of $\phi(r)$ as
$$
\theta_1=\int_0^1 r \phi(r) dr;~~\theta_2 = \int_0^1 r^2 \phi(r) dr,
$$ respectively. Note that, if  the probability density  $\phi(r)$
is not symmetric around $r=1/2$, it can be shown that, in the
hydrodynamic  equation for density field, drift dominates
diffusion unless the asymmetry is small and comparable to the
diffusive contribution. In that case, the analysis would lead to
hyperbolic hydrodynamic equations for density field (hydrodynamics of such systems are discussed in Sec. \ref{AMCM}). Here we consider the density function $\phi(r)$, which is symmetric
around $r=1/2$, i. e., $\phi(r) = \phi(1-r)$, thus $\theta_1
={1}/{2}$ is taken throughout; but the probability  density
$\phi(r)$ is otherwise arbitrary.

{\it Breakdown of Kolmogorov criterion.}-- In this model with random sequential update, at any instant of time, mass is chipped off from a single departure site and then it arrives at its two nearest-neighbor destination sites. Clearly, the reverse path, where mass would have been simultaneously chipped from two departure sites $i-1$ and $i+1$ and would have arrived at a single destination site $i$, is not allowed by the actual dynamics as given in eq.(\ref{model-1}). Therefore, Kolmogorov criterion is violated and consequently there is no detailed balance even when there is as such no external biasing force.

{\it Dynamics when $F \ne 0$.} -- Let us now bias the system by applying a small constant biasing force field
$\vec{F} = F \hat{x}$, say, along the clockwise direction, which
affects the mass transfer rates according to eq. (\ref{bias2}). Since, at
every instant of time, two fractions of the chipped-off mass from
site $i$ could be simultaneously transferred, to the two
neighboring sites $i\pm 1$, the modified biased rate in this case
is written as \be c^F_{i \rightarrow \{ i+1, i-1\} } = c_{i
\rightarrow \{ i+1, i-1\} } \left[1 + \frac{\Delta e_i}{2}\right],
\label{rate-model-I-RSU} \ee where $c_{i \rightarrow \{i+1, i-1\}
} = 1$ and $\Delta e_i = Fa (\Delta m_{i \rightarrow i+1} - \Delta
m_{i \rightarrow i-1})$ with $\Delta m_{i \rightarrow i+1}=\tilde
\lambda r_i m_i$ and $\Delta m_{i \rightarrow i-1}=\tilde \lambda
(1-r_i) m_i$. For notational simplicity, we denote the biased rate
as $c^F_{i \rightarrow \{ i+1, i-1\} } \equiv c_i^F$, which can be
explicitly written as $c_i^F = 1 + \tilde{\lambda} (2r_i-1) m_i
{Fa}/{2},$ with $\tilde{\lambda} = 1 - \lambda$. We now write the
modified dynamics, \be m_i(t+dt) =
\begin{cases}
\lambda m_i(t),~~~\mbox{prob.}~ c_i^Fdt \\
m_i(t)+ \tilde{\lambda}r_{i-1}m_{i-1}(t),  ~~\mbox{prob.} ~c_{i-1}^F dt\\
m_i(t)+\tilde{\lambda}\tilde{r}_{i+1}m_{i+1}(t), ~~\mbox{prob.} ~c_{i+1}^F dt\\
m_i(t), ~~ \mbox{prob.} ~(1-(c_i^F+c_{i-1}^F+c_{i+1}^F)dt).
\end{cases}
\ee Consequently the time  evolution of the first moment of mass
$m_i(t)$ in the infinitesimal time $dt$ can be written as \bea
\langle m_i(t+dt) \rangle &=& \langle \lambda m_i(t) c^F_i \rangle
dt
\nonumber \\
&+& \langle [m_i(t)
+ \tilde \lambda r_{i-1} m_{i-1}(t)] c^F_{i-1} \rangle dt
\nonumber \\
&+& \langle [m_i(t)+ \tilde \lambda \tilde{r}_{i+1}m_{i+1}(t)] c^F_{i+1}
\rangle dt
\nonumber \\
&+& \langle m_i(t) [1-(c_i^F+c_{i-1}^F+c_{i+1}^F)dt] \rangle
\nonumber \eea  After  simplifying the above expression, the time
evolution of average mass, or mass density, $\langle m_i \rangle
\equiv \rho_i$ at site $i$, can be rewritten as
\begin{eqnarray}
\frac{d\rho_i}{dt}&=&\tilde{\lambda}\langle r_{i-1} m_{i-1} c_{i-1}^F + (1-r_{i+1}) m_{i+1} c_{i+1}^F - m_i c_i^F \rangle
\nonumber \\
&=& \frac{\tilde{\lambda}}{2} (\rho_{i-1} + \rho_{i+1} -2 \rho_i)
\nonumber \\
&& + \frac{\tilde{\lambda}^2}{2}(2\theta_2 - 1/2) [\langle m_{i-1}^2 \rangle Fa - \langle m_{i+1}^2 \rangle Fa].
\label{density_model1}
\end{eqnarray}
Note that the time evolution of the
first moment of local mass, i.e., the density $\rho_i = \langle
m_i \rangle$, depends on the second moments $\langle m_{i \pm 1}^2
\rangle$ of neighboring masses, and so on. Thus the hierarchy between 
the local density and the local fluctuation does not close.

{\it Hydrodynamics}-- However, we are interested in the 
hydrodynamic  description of the density field at large space and 
time scales, called diffusive scaling limit as described below. Importantly, on the large spatio-temporal scales, local observables are expected to be 
slowly varying functions of space and time. Therefore, we could safely assume that a local steady state is
achieved throughout the system such that average of any local
observable $g(m_i)$ could be replaced by its exact local
steady-state average $\langle g(m_i) \rangle_{st}$, which in that
case would be a function of the local density $\rho_i$ only. In
other words, we assume $\langle g(m_i) \rangle \approx \langle
g(m_i) \rangle_{st}$. Thus, for the average of the quantity
$g(m_i)=m_i^2$, i.e., the second moment of local mass, we have
replaced the average by the its local steady-state average, \be
\langle m_i^2 \rangle \approx \langle m_i^2 \rangle_{st} = \frac{1}{1 - 2 \tilde \lambda \theta_2} \rho_i^2.
\label{2nd-moment-model1} \ee 
The above steady-state expression of
the second moment has been exactly calculated before in Ref.
\cite{Das_PRE2016}. Now substituting eq. \ref{2nd-moment-model1}
in eq. (\ref{density_model1}) and then taking the diffusive
scaling limit of eq. (\ref{density_model1}), $i \rightarrow x =
i/L$, $t \rightarrow \tau = t/L^2$ and $a \rightarrow 1/L$, we
obtain the hydrodynamic equation for the density field, $\partial_{\tau} \rho(x, \tau) + \partial_x J=0$,
where current $J(\rho(x, \tau))$ is given by
\begin{equation}
J = \frac{\tilde{\lambda}^2}{2} \frac{4\theta_2-1}{1-2 \tilde{\lambda} \theta_2} \rho^2 F - \frac{\tilde{\lambda}}{2} \frac{\partial \rho}{\partial x}.
\end{equation}
In the above equation, we break the current $J = J_d + J_D$ into
two parts, drift current $J_d = [{\tilde{\lambda}^2 (4 \theta_2 -
1)}/{2(1-2 \tilde{\lambda} \theta_2)}] \rho^2 F$ and diffusive
current $J_D = - ({\tilde{\lambda}}/{2}) ({\partial
\rho}/{\partial x})$, to identify the conductivity and the
diffusion coefficient, respectively, as \bea \chi(\rho) =
\frac{\tilde{\lambda}^2}{2} \frac{(4 \theta_2 - 1)}{(1 - 2
\tilde{\lambda} \theta_2)} \rho^2 ~;~ D(\rho) =
\frac{\tilde{\lambda}}{2}. \label{D-model1} \eea Now the scaled
variance $\sigma^2(\rho)$ of sub-system mass  [as defined in eq.
(\ref{sigma})] can be calculated by summing over the microscopic
correlation function $c(n) = \langle m_i m_{i+n} \rangle -
\rho^2$, \be \sigma^2(\rho) = \sum_{n = -\infty}^{\infty} c(n) =
\frac{\tilde \lambda (4 \theta_2 - 1)}{(1 - 2 \tilde \lambda
\theta_2)} \rho^2 \equiv \frac{\rho^2}{\eta}, \label{sigma-model1}
\ee where $c(n)$ has been exactly calculated in Ref.
\cite{Das_PRE2016},
\begin{eqnarray}
c(n) &=& \frac{2 \tilde \lambda \theta_2}{(1 - 2 \tilde \lambda \theta_2)} \rho^2 \mbox{~~~~~~~~~~for $n=0$} \nonumber \\
&=& - \frac{\tilde \lambda}{2} \frac{(1 - 2 \theta_2)}{(1 - 2 \tilde \lambda \theta_2)} \rho^2 \mbox{~~~~~for $n=1$} \nonumber \\
&=& 0 \mbox{~~~~~~~~~~~~~~~~~~~~~~~~~~for $n\geq 2$,} \nonumber
\end{eqnarray}
and $\eta = {(1 - 2 \tilde \lambda \theta_2)}/{\tilde \lambda (4 
\theta_2 - 1)}$. 
Using eqs. (\ref{D-model1}) and (\ref{sigma-model1}), 
one can readily verify that the ER as in eq. 
(\ref{ER}) is indeed satisfied. We emphasize that the nearest-
neighbor spatial correlations here (also in the other models 
discussed later) are actually finite and our hydrodynamic analysis 
takes into account the effects of the finite microscopic spatial 
correlations.

\subsubsection{Parallel Update}
\label{model-I-PU}

In Model I with parallel update (PU), fractions of masses  to be
transferred to the two nearest neighbors are the same as in the
case of random sequential update. However, at a discreet time $t$,
the mass variables at all sites are updated simultaneously
according to the following rule, \be m_i(t+1) = \lambda m_i(t) +
\tilde \lambda r_{i-1}m_{i-1}(t) + \tilde \lambda
\tilde{r}_{i+1} m_{i+1}(t), \label{mcm-I-PU} \ee where $\tilde \lambda = 1 -
\lambda$, $\tilde r_i = 1-r_i$ and $r_i \in ({0,1})$ is a symmetrically distributed
random variable, having a probability density $\phi(r_i)$. The
time evolution equation in the configuration space $\{m_i \}
\equiv \{m_1, m_2, \dots, m_L\}$ can be written as \bea {\cal
P}[\{m_i \}, t+1] = \left[ \prod_j \int d m_j
\right]~~~~~~~~~~~~~~~~~~~~~~
\nonumber \\
~~~~~~~~~~~~\times \Gamma[\{m_j \} \rightarrow \{m_i \}] {\cal
P}[\{m_j \}, t], \label{P_C_t} \eea where ${\cal P}[\{m_i \}, t]$
is the probability of a  configuration $\{m_i \}$ at time $t$ and
$$
\Gamma[\{m_j \} \rightarrow \{m_i \}] = \prod_i \phi(r_i)
$$
is the transition probability, per unit time, from a
configuration $\{m_j \}$ to another configuration $\{m_i \}$.

{\it Breakdown of Kolmogorov criterion.}-- In the case of parallel update, the breakdown of Kolmogorov criterion, though not quite obvious, can be straightforwardly shown for generic parameter values $\lambda \ne 0$. 
For example, consider a configuration having two sites $i-1$ and $i$, with masses $m_{i-1}$ finite and $m_i$ infinitesimal (say, $m_i=0$, just for the sake of argument), respectively. Then, a chunk of mass is transferred from site $(i-1)$ to site $i$ so that $m_{i-1} \rightarrow m_{i-1}'>0$ and $m_i \rightarrow m_i'>0$. In the next time step, since at least a $\lambda$ fraction of mass $m_i'$ must be retained at site $i$, the reverse path where the whole mass $m_i'$ would have been transferred back to $i-1$ from site $i$ is not possible, implying breakdown of Kolmogorov criterion and thus violation of detailed balance. This simple, though not rigorous, argument can be readily extended to any configuration with sufficiently large difference of masses in any two neighbouring sites so that there cannot be a reverse path corresponding to a particular possible path of mass transfer. Note that, in this argument, we consider only the unbiased process ($F=0$). Let us consider transitions $\{m_i\} \rightarrow \{m_i'\}$ and $\{m_i'\} \rightarrow \{m_i''\}$ at two consecutive time steps. In the second transition, one must have $m_i'' > \lambda m_i'$, i.e., the mass retained at site $i$ must be at least $\lambda m_i'$. Now, if the amount of mass $\lambda m_i'$ is greater than mass $m_i$, the value of mass at site $i$ at the initial step, clearly the path cannot be reversed. Therefore the condition for which a process cannot be reversed is simply $\lambda m_i' > m_i$, which, after using eq. (\ref{mcm-I-PU}) $m_i'=\lambda m_i + \tilde{\lambda}r_{i-1} m_{i-1} + \tilde{\lambda} \tilde{r}_{i+1} m_{i+1}$, leads to the condition
\begin{equation}
r_{i-1}m_{i-1} + \tilde{r}_{i+1}m_{i+1} - \frac{1+\lambda}{\lambda}m_i>0.
\end{equation}
Therefore, for $\lambda \ne 0$, indeed there are configurations (a finite set in the configuration space) which satisfy the above inequality. This implies breakdown of Kolmogorov criterion and that the steady state is far from equilibrium even in the absence of any biasing force ($F=0$). Analysis for $\lambda=0$ requires more effort and is omitted here.

{\it Dynamics when $F \ne 0$.} --
Let us now consider the  process in the presence of an externally applied biasing force, $F \ne 0$. Once the random fractions ($\tilde \lambda r_i m_i$ and  $\tilde
\lambda (1-r_i) m_i$)  of mass $m_i$ at a site $i$ are chosen at
time $t$ they  are transferred, at the next discrete time step, to
the nearest neighbour sites $i+1$ and $i-1$, respectively, with
probability $1$ and this is done simultaneously for all sites.
That is, in this case, the mass transfer rate, or the transition
probability per unit time, can be written as $ c_{\{i_n\} \rightarrow
\{j_{n'}\} } = 1$, which we modify, in the presence of biasing
force, as $c^F_{\{i_n\} \rightarrow \{j_{n'}\} } = c_{\{i_n\}
\rightarrow \{j_{n'}\} } \prod_i \exp(\Delta e_i/2)$, according to
eq. (\ref{bias2}). Here we put $\Phi(\Delta e) = \exp(\Delta
e/2)$ with $\Delta e = \sum_i \Delta e_i$ and $\Delta e_i = Fa
(\Delta m_{i \rightarrow i+1} - \Delta m_{i \rightarrow i-1}) =
\tilde{\lambda} (2r_i-1) m_i Fa$. The time evolution eq.
(\ref{P_C_t}) can now be written by replacing the original
transition probability $\Gamma$ with the modified one $\Gamma^F$,
\bea \Gamma^F[\{m_j \} \rightarrow \{m_i \}] = \prod_j \left[
\frac{\phi(r_j) e^{\Delta e_j/2}}{\gamma(m_j, F)} \right],
\label{bias3} \eea where $\gamma(m_j, F)$ is a normalization
constant, ensuring that the transition probability $\Gamma^F[.]$
is suitably assigned from a normalized probability density
function where $(\prod_j \int dr_j) \Gamma^F = 1$. As the
probability density $\phi(r)$ is considered to be symmetric about
$r=1/2$, we have the following expansion in powers of $F$, \bea
\gamma (m_i, F) &=& \int_0^1 \phi(r_i) e^{\tilde{\lambda} (2r_i-1)
m_i Fa/2} dr_i
\nonumber \\
&=& 1 + \frac{(\tilde \lambda m_i)^2 \theta_2}{8} (Fa)^2 + \dots ,
\nonumber \eea implying that the leading order term is quadratic
${\cal O}(F^2)$ in  the biasing force $F$ and therefore, to linear
order of $F$, we can take {$\gamma(m_i, F) \approx 1$} in the
following analysis (see also Sec. \ref{Model-II-PU}).

The expression for the average of mass $m_i$ at site $i$ can now
be  written as \bea && \langle m_i(t+1) \rangle = \left[ \prod_j
\int dm_j \right] m_i {\cal P}[\{m_j\},t+1]
\nonumber \\
&=& \left\langle \left[ \lambda m_i(t)+ \tilde \lambda
r_{i-1}m_{i-1}(t) + \tilde \lambda (1-r_{i+1})m_{i+1}(t) \right]
\right\rangle,  \nonumber \eea where the angular brackets $\langle
\cdot \rangle$ denote average over  both random numbers $\{r_j\}$ and
the mass variables $\{m_j\}$. Explicitly writing the terms, we get
\bea \langle m_i(t+1) \rangle = \left\langle \lambda m_i(t) \int
\frac{\phi(r) e^{\tilde \lambda (2r-1)m_i Fa/2}}{\gamma(m_i, F)}
dr \right\rangle
\nonumber \\
+ \left\langle \tilde \lambda m_{i-1}(t) \int r \frac{\phi(r) e^{\tilde \lambda (2r-1) m_{i-1} Fa/2}}{\gamma(m_{i-1}, F)} dr \right\rangle
\nonumber \\
+ \left\langle \tilde \lambda m_{i+1}(t) \int (1-r) \frac{\phi(r) e^{\tilde \lambda (2r-1) m_{i+1} Fa/2}}{\gamma(m_{i+1}, F)} dr \right\rangle, \nonumber 
\eea
which, in leading order in $F$, leads to
\begin{eqnarray}
\rho_i(t+1) -\rho_i(t) = \frac{\tilde{\lambda}}{2}(\rho_{i-1} +\rho_{i+1}-2\rho_i)\nonumber\\
+\frac{\tilde{\lambda}^2}{2}(2\theta_2 -1/2)[\langle m_{i-1}^2\rangle Fa - \langle m_{i+1}^2\rangle  Fa]. \nonumber
\end{eqnarray}

{\it Hydrodynamics.}-- Now taking the diffusive scaling limit $i  \rightarrow x=i/L$,  $t
\rightarrow \tau=t/L^2$ and $a \rightarrow 1/L$ and substituting
$\langle m_i^2 \rangle$ by the following expression of second moment 
 \cite{Das_PRE2016}, within the assumption of local steady-state,
$$
\langle m_i^2 \rangle_{st} = \frac{1}{\epsilon + (1-\epsilon) \sqrt{\frac{\kappa-1}{\kappa+1} }} \rho_i^2,
$$
we obtain the hydrodynamic equation for the density field,
${\partial_{\tau} \rho(x, \tau)} + {\partial_x} (J_d + J_D) = 0$,
where the drift current $J_d(\rho(x, \tau))$ and the diffusive current $J_D(\rho(x, \tau))$ are
given by \bea J_d =
\frac{\tilde{\lambda}^2}{2}\frac{4\theta_2-1}{\epsilon  +
(1-\epsilon) \sqrt{\frac{\kappa-1}{\kappa+1} }} \rho^2 F ~;~ J_D =
-\frac{\tilde{\lambda}}{2} \frac{\partial \rho}{\partial x}, \eea
respectively. Then, the conductivity $\chi(\rho)$ and the diffusion 
coefficient $D(\rho)$ can be expressed as \bea \chi(\rho) =
\frac{\tilde{\lambda}^2}{2} \frac{4\theta_2-1}
{\epsilon+(1-\epsilon)\sqrt{\frac{\kappa-1}{\kappa+1}}}\rho^2 ~;~
D(\rho) = \frac{\tilde{\lambda}}{2}, \eea where $\epsilon = 2 - 4
\theta_2$ and $\kappa = (1+\lambda)/(1-\lambda)$. The Einstein
relation eq. (\ref{ER}) can be immediately verified using the 
expression of the scaled variance,
$$
\sigma^2(\rho) = \frac{\tilde \lambda (4\theta_2-1)}{\epsilon+(1-\epsilon)\sqrt{\frac{\kappa-1}{\kappa+1}}}\rho^2,
$$
which was exactly calculated earlier in Ref. \cite{Das_PRE2016}. We
mention here that the microscopic spatial correlations, as in the
case of Model I (RSU), are also finite and have been accounted
exactly in the above analysis.

\subsection{Model II}
\label{Model-II}

The class of models studied in this section is a generalized version 
of previously known Hammersley process \cite{Aldous} and a variant of 
random average processes \cite{Ferrari}. These models were studied in 
the past to understand force fluctuations in granular beads
\cite{Majumdar_Science1995, Majumdar_PRE1996} and dynamics of
driven interacting particles on a ring \cite{Krug_JSP2000,
Zielen_JSP2002}, etc.

\subsubsection{Random Sequential Update}

In Model II with random sequential update,  two nearest neighbor
sites are updated in an infinitesimal time $dt$: A random fraction
of mass at site $i$ is chipped off and transferred either to site
$i-1$ or to site $i+1$, each with probability $(1/2) dt$, i.e.,
the mass transfer rates $c_{i \rightarrow i-1}=1/2$ and $c_{i
\rightarrow i+1}=1/2$.

{\it Breakdown of Kolmogorov criterion.}-- First let us show that the process in the absence of any external bias ($F=0$) violates Kolmogorov criterion and therefore also detailed balance. 
Let us consider transitions $\{m_i\} \rightarrow \{m_i'\}$ and $\{m_i'\} \rightarrow \{m_i''\}$ at two consecutive time steps where
\begin{eqnarray}
m'_i = (1-\tilde{\lambda}r_i) m_i ~;~ m'_{i+1} = m_{i+1}+\tilde{\lambda}r_i m_i, \nonumber
\\
m''_i = (1-\tilde{\lambda}r_i) m_i' ~;~ m''_{i+1} = m_{i+1}' + \tilde{\lambda} r_i m_i', \nonumber
\end{eqnarray}
Now the conditions, $m_i''=m_i$ and $m_{i+1}''=m_{i+1}$, for the existence of a reverse path leads to an equality,
$r'_{i+1}={r_i m_i}/{(m_{i+1} + \tilde{\lambda}r_i m_i)}$. Or equivalently, an inequality $m_{i+1} \ge \lambda r_i m_i$, as $r'_{i+1} \le 1$, must be satisfied for the existence of a reverse path. Said differently, the condition for which a reverse path will  not exist can be written as the following inequality on the ratio of neighboring masses,
\begin{equation}
\frac{m_i}{ m_{i+1}} > \frac{1}{\lambda r_i}. \nonumber
\end{equation}
The above condition is satisfied by a finite set in the configuration space and will then imply the steady state to be far from equilibrium even in the absence of any external biasing force ($F=0$).

{\it Dynamics when $F \ne 0$.} -- However, in the presence of a biasing force $F \ne 0$, the dynamics is modified as
\be m_i(t+dt) = 
\begin{cases}
\lambda m_i(t)+\tilde{\lambda}(1-r_i)m_i(t) ~~~~~~ \mbox{prob.} ~dt & \\
m_i(t) + \tilde \lambda r_{i+1} m_{i+1}(t) ~~~  \mbox{prob.} ~c_{i+1 \rightarrow i}^Fdt &\\
m_i(t) + \tilde \lambda r_{i-1}m_{i-1}(t)~~~ \mbox{prob.}  ~c_{i-1 \rightarrow i}^Fdt &\\
m_i(t) ~~~ \mbox{prob.}~ [1-(1+c_{i+1 \rightarrow i}^F + c_{i-1 \rightarrow i}^F)dt]&
\end{cases}
\ee
where $\tilde{\lambda}=1-\lambda$ and the modified  mass transfer
rates, $c^F_{i \rightarrow i \pm 1} = \exp(\pm \Delta m_{i
\rightarrow i\pm 1} Fa/2)$ with transported mass $\Delta m_{i
\rightarrow i \pm 1} = \tilde{\lambda}r_im_i(t)$, can be written,
in leading order of $F$, as
\begin{eqnarray}
c_{i-1 \rightarrow i}=\frac{1}{2} + \frac{\tilde{\lambda}}{4}r_{i-1} m_{i-1} F a,
\label{rate1-model-II-RSU} \\
c_{i+1 \rightarrow i}=\frac{1}{2} - \frac{\tilde{\lambda}}{4} r_{i+1} m_{i+1} F a.
\label{rate2-model-II-RSU}
\end{eqnarray}
Clearly, $F=0$ reproduces the original unbiased dynamics. Now,
the time evolution of average mass or density at site $i$ is given
by,
\begin{eqnarray}
\frac{d\langle m_i\rangle}{dt} &=&\tilde{\lambda}\langle [ r_{i-1}m_{i-1}c_{i-1 \rightarrow i}^F+r_{i+1} m_{i+1} c_{i+1 \rightarrow i}^F-r_im_ic_i^F ] \rangle
\nonumber
\end{eqnarray}
which, in leading order of $F$, can be written as
\begin{eqnarray}
\frac{d \rho_i}{dt}&=&\frac{\tilde{\lambda}}{2}\theta_1(\rho_{i-1}+\rho_{i+1}-2\rho_i)\nonumber\\
&&+\frac{\tilde{\lambda}^2}{4}\theta_2[\langle m_{i-1}^2\rangle Fa - \langle m_{i+1}^2\rangle Fa]. \nonumber
\end{eqnarray}

{\it Hydrodynamics}-- Taking the diffusive scaling  limit $i\rightarrow x=i/L$,
$t\rightarrow \tau=t/L^2$ and $a \rightarrow 1/L$ and using the
local steady-state expression for the second moment,
$$
\langle m_i^2 \rangle = \frac{\theta_1}{\theta_1 - \tilde{\lambda} \theta_2} \rho_i^2,
$$
we obtain the hydrodynamic equation governing  the density field,
$ {\partial_{\tau} \rho(x, \tau)} + {\partial_x} (J_d + J_D) =0,$
where the drift current $J_d(\rho(x, \tau))$ and the diffusive  current $J_D(\rho(x, \tau))$ are
given by \bea J_d = \frac{\tilde{\lambda}^2}{2}
\frac{\theta_1\theta_2}{\theta_1 - \tilde{\lambda}\theta_2} \rho^2 F 
~;~J_D = -\frac{\tilde{\lambda}}{2} \theta_1 \frac{\partial
\rho}{\partial x}. \eea Therefore, the conductivity $\chi(\rho)$ and
the diffusion coefficient $D(\rho)$ are given by \bea \chi(\rho) =
\frac{\tilde{\lambda}^2}{2}\frac{\theta_1\theta_2}{\theta_1 - \tilde{\lambda}\theta_2}\rho^2
~;~ D(\rho) = \frac{\tilde{\lambda}}{2} \theta_1. \eea The
Einstein relation eq. (\ref{ER}) can now be verified by using the expression of scaled variance,
$$
\sigma^2(\rho) =  \frac{\tilde \lambda \theta_2}{\theta_1 - \tilde \lambda \theta_2} \rho^2,
$$
which was obtained earlier in Ref. \cite{Das_PRE2016}.

\subsubsection{Parallel Update}
\label{Model-II-PU}

\noindent

In Model II with parallel update, at each discrete  time step,
masses at all sites are updated simultaneously according to the
following rule,
\begin{eqnarray}
m_i(t+1)
&=& (1- \tilde \lambda r_i) m_i(t)+ \tilde \lambda r_{i+1} m_{i+1}(t) \nonumber \\
&+& \tilde \lambda [ s_{i-1} r_{i-1} m_{i-1}(t) -s_{i+1} r_{i+1} m_{i+1}(t) ]~~~ 
\label{model2-PU}
\end{eqnarray}
where $\tilde \lambda = 1-\lambda$. Here we have  introduced a set
of discrete i.i.d. random variables $\{ s_i \}$: When the
chipped-off fraction of mass moves to right, $s_i=1$ and otherwise
$s_i=0$. As each of the values $s_i=0$ and $s_i=1$ occurs with
probability ${1}/{2}$, we have $\langle s_i^n\rangle = {1}/{2}$
for $n > 0$.

{\it Breakdown of Kolmogorov criterion.}-- In this model, the breakdown of Kolmogorov criterion, for generic parameter values $\lambda \ne 0$, can be shown along the lines of arguments as given in the case of parallel update for model I in Sec. \ref{model-I-PU}. 
As before, let us consider transitions $\{m_i\} \rightarrow \{m_i'\}$ and $\{m_i'\} \rightarrow \{m_i''\}$ at two consecutive time steps.
Provided that the mass $(1-\tilde{\lambda}r'_i) m'_i$, the least amount of mass retained at site $i$ after second transition, is greater than the initial mass $m_i$, there cannot be a reverse path. Using dynamical rule in eq. (\ref{model2-PU}), it can be shown that the condition of inequality $(1-\tilde{\lambda}r'_i) m'_i > m_i$ leads to a condition on the initial masses,
\begin{equation}
s_{i-1}r_{i-1}m_{i-1}+(1-s_{i+1})r_{i+1}m_{i+1}-\left(\frac{1}{\lambda}+r_i\right)m_i>0. \nonumber
\end{equation}
The above condition is satisfied for a finite set of configurations in the configuration space and will then imply violation of Kolmogorov criterion, and thus also detailed balance, and that the steady state is far from equilibrium even in the absence of any biasing force ($F=0$).

{\it Dynamics when $F \ne 0$.} -- In the presence of a biasing force $F \ne 0$, the transition probability $\Gamma[\{m_j\}
\rightarrow \{m_k\}]$ from a configuration $\{m_j\}$ to another
configuration $\{m_k\}$ is modified as
\begin{equation}
\Gamma^F[\lbrace m_j \rbrace \rightarrow \lbrace m_k \rbrace] = \prod_i \left[\frac{1}{\gamma (m_i, F)} \phi(r_i) e^{\Delta e_i/2} \right],
\end{equation}
where $\Delta e_i = [s_i-(1-s_i)] \tilde \lambda r_i m_i Fa$ and
the normalization factor \bea \gamma (m_i, F) &=& \Sigma_{s_i}
P(s_i) \int_0^1 \phi(r_i) e^{(2s_i-1) \tilde \lambda r_i m_i Fa/2}
dr_i
\nonumber \\
&=& 1 + \frac{(\tilde{\lambda}m)^2 \theta_2}{8} (Fa)^2 + \dots
\approx 1, \nonumber \eea to the linear order of $F$. The time  evolution of the average mass or density at
site $i$ is given by
\begin{eqnarray}
\left\langle m_i(t+1) \right\rangle &=& \left\langle (1 - \tilde{\lambda} r_i) m_i(t) \right\rangle + \left\langle \tilde{\lambda} s_{i-1} r_{i-1} m_{i-1}(t) \right\rangle \nonumber \\
&&~~~~~~+ \left\langle \tilde{\lambda} (1-s_{i+1}) r_{i+1}
m_{i+1}(t) \right\rangle \eea where the above angular  brackets
denote averaging over all three random variables, $\{r_i\}$,
$\{s_i\}$ and $\{m_i\}$. Equivalently, we can write \bea &&
\left\langle m_i(t+1) \right\rangle = \left\langle (1 -
\tilde{\lambda} r_i) m_i(t) \int \frac{\phi(r_i) e^{\Delta
e_i/2}}{\gamma(m_i, F)}dr_i \right\rangle
\nonumber \\
&&+ \left\langle \tilde{\lambda} s_{i-1} r_{i-1} m_{i-1}(t) \int \frac{\phi(r_{i-1}) e^{\Delta e_{i-1}/2}}{\gamma(m_{i-1}, F)} dr_{i-1} \right\rangle \\
&&+ \left\langle \tilde{\lambda} (1-s_{i+1}) r_{i+1} m_{i+1}(t) \int \frac{\phi(r_{i+1}) e^{\Delta e_{i+1}/2}}{\gamma(m_{i+1}, F)} dr_{i+1} \right\rangle, \nonumber
\end{eqnarray}
where, in the second step, we have explicitly written the
averaging over the i.i.d. random variables $\{r_i\}$. Next, doing
the averaging over the i.i.d. random variables $\{s_i\}$, we
obtain, in linear order of $F$, the time evolution equation for
density $\rho_i=\langle m_i \rangle$ at site $i$,
\begin{eqnarray}
\rho_i(t+1)-\rho_i(t) &=&\frac{\tilde{\lambda}}{2}\theta_1(\rho_{i-1}+\rho_{i+1}-2\rho_i)
\nonumber\\
&+& \frac{\tilde{\lambda}^2}{4} \theta_2 [\langle m_{i-1}^2 \rangle Fa - \langle m_{i+1}^2 \rangle Fa].~~~
\end{eqnarray}

{\it Hydrodynamics.}-- Now rescaling the space and time by $i \rightarrow x=i/L$, $t
\rightarrow \tau = t/L^2$ and $a \rightarrow 1/L$, and using the 
expression for second moment of $m_i$ in the local steady state \cite{Das_PRE2016},
$$
\langle m_i^2 \rangle = \frac{\sqrt{\alpha}}{1-(1-\lambda)\epsilon} \rho_i^2,
$$
we obtain the hydrodynamic equation of density field, $\partial_{\tau} \rho(x, \tau) + {\partial_x} (J_d + J_D) = 0$
where the drift $J_d(\rho(x, \tau))$ and diffusive currents $J_D(\rho(x, \tau))$, respectively, can be written as
\begin{equation}
J_D = -\frac{\tilde{\lambda}}{2}\theta_1\frac{\partial \rho}{\partial x}~;~J_d = \frac{\tilde{\lambda}^2}{2} \theta_2 \frac{\sqrt{\alpha}}{1-(1-\lambda) \epsilon} \rho^2 F.
\nonumber
\end{equation}
The above expressions of  currents immediately gives the diffusion
coefficients and the conductivity as a function of density, \be
\chi(\rho) = \frac{\tilde{\lambda}^2}{2}\theta_2
\frac{\sqrt{\alpha}}{1-(1-\lambda) \epsilon} \rho^2 ~;~ D(\rho) =
\frac{\tilde{\lambda}}{2} \theta_1, \ee respectively, with
$\epsilon = \theta_2/\theta_1$, $\alpha = {(1 + \lambda)}/{2}$.  Now, by
using the exact expression of scaled variance \cite{Das_PRE2016},
\begin{equation}
\sigma^2(\rho) = \frac{\tilde{\lambda} \sqrt{\alpha}\epsilon}{1-\tilde{\lambda} \epsilon},
\nonumber
\end{equation}
one can verify that the  ER as in eq. (\ref{ER}) is
indeed satisfied. Note that, as in the case of Model I, the
microscopic spatial correlations are also finite here and have
been taken into account in deriving hydrodynamics.

\subsection{Model III}
\label{Model-III}

This class of models have been studied intensively  in the past to
understand distribution of wealth in a population \cite{wealth_review,wealth,
Patriarca_EPJB2010}. In this model, each site keeps a $\lambda$
fraction (usually called ``saving propensity'' in the literature)
of its own mass, and the remaining mass of two neighboring sites
are mixed and are distributed randomly among themselves. Here we
study only the random sequential update dynamics, which can be
written in an infinitesimal time $dt$ as follows,
\bea && m_i(t+dt) \nonumber \\ 
&& = \begin{cases}
\lambda m_i(t) + \tilde \lambda r_i [m_i(t) + m_{i+1}(t)] ~~~~~\mbox{prob.}~c_{i+1 \rightarrow i}dt \\
\lambda m_i(t) + \tilde \lambda \tilde r_{i-1} [m_i(t)+m_{i-1}(t)] ~~\mbox{prob.}~c_{i-1 \rightarrow i}dt \\
m_i(t) ~~~~~~~~~~~~~~~~\mbox{prob.}~ [1-(c_{i+1 \rightarrow i} + c_{i-1 \rightarrow i})dt]
\end{cases}\nonumber\\
\eea
where $\tilde \lambda = 1-\lambda$, $\tilde r_i = 1- r_i$, $m_i(t)$ is mass  at site $i$
at time $t$, $r_i \in (0,1)$ is a i.i.d. random variable having a
probability density $\phi(r_i)$ (symmetric around $1/2$) and the
mass transfer rate $c_{i \rightarrow j}=1$ (here $j=i\pm 1$).

{\it Violation of Kolmogorov criterion.}-- Again, let us consider transitions $\{m_i\} \rightarrow \{m_i'\}$ and $\{m_i'\} \rightarrow \{m_i''\}$ at two consecutive time steps where, by denoting $\mu_{i,i+1}=m_i+m_{i+1}$,
\begin{eqnarray}
m'_i = \lambda m_i +\tilde{\lambda} r_i\mu_{i,i+1} ~;~ m'_{i+1}= \lambda m_{i+1} +\tilde{\lambda} \tilde{r}_i\mu_{i,i+1}, \nonumber
\\
m''_i = \lambda m_i' +\tilde{\lambda} r_i \mu_{i,i+1} ~;~ m''_{i+1}= \lambda m'_{i+1} +\tilde{\lambda} \tilde{r}_i \mu_{i,i+1}. \nonumber
\end{eqnarray}
The condition, $m''_i = m_i$ and $m''_{i+1} = m_{i+1}$, of having a reverse path can be written as an equality $r'_{i} = {(1+\lambda)m_i}/{\mu_{i,i+1}}-\lambda r_i$, or alternatively, as 
an inequality (as $r'_{i} \le 1$) on the ratio of neighboring masses $m_i/m_{i+1} \le (1+\lambda r_i)/\lambda \tilde r_i$. Said differently, for $m_i/m_{i+1} > (1+\lambda r_i)/\lambda \tilde r_i$, Kolmogorov criterion and detailed balance are violated, and thus the steady state is far away from equilibrium even in the absence of any biasing force ($F=0$).

{\it Dynamics when $F \ne 0$.} --
In the presence of a biasing force, the mass transfer rates are
modified as \bea c^F_{i+1 \rightarrow i} &=& e^{-\Delta m_{i+1
\rightarrow i} F a/2} \approx 1 - \frac{1}{2} \Delta m_{i+1
\rightarrow i} F a,
\nonumber \\
c^F_{i-1 \rightarrow i} &=& e^{\Delta m_{i-1 \rightarrow i} F a/2}
\approx 1 + \frac{1}{2} \Delta m_{i-1 \rightarrow i} F a \nonumber
\eea where \bea \Delta m_{i+1 \rightarrow i} &=& \tilde \lambda
r_i m_{i+1}(t) - \tilde \lambda (1-r_i) m_i(t) \label{delta-m1}
\eea  and \bea \Delta m_{i-1 \rightarrow i} &=& \tilde \lambda
(1-r_{i-1}) m_{i-1} - \tilde \lambda r_{i-1} m_i(t).
\label{delta-m2} \eea The time evolution of the first moment of
local mass or  density  $\rho_i = \langle m_i \rangle$ at site $i$
can be written as \bea \langle m_i(t+dt)\rangle = \left\langle
[\lambda m_i(t)+\tilde{\lambda}r_i(m_i(t)+m_{i+1}(t))] c_{i+1
\rightarrow i }^Fdt \right\rangle
\nonumber \\
+ \left\langle [\lambda m_i(t)+\tilde{\lambda}(1-r_{i-1})(m_i(t) +
m_{i-1}(t))] c_{i-1 \rightarrow i}^Fdt \right\rangle
\nonumber \\
+ \left\langle m_i(t) [1 - (c_{i-1 \rightarrow i}^F + c_{i+1
\rightarrow i}^F) dt] \right \rangle. \nonumber \eea After
substituting  $\langle m_i \rangle = \rho_i$ and some
simplifications, we have the following evolution for density
$\rho_i$, \bea  \frac{d\rho_i}{dt}  &=& \frac{\tilde{\lambda}}{2}[
\rho_{i+1}-2\rho_i +\rho_{i-1}]
\nonumber \\
&-& \frac{1}{2}\left\langle (\Delta m_{i+1 \rightarrow i}^2Fa - \Delta m_{i-1 \rightarrow i}^2 Fa) \right \rangle, \nonumber
\eea
which leads to
\bea
&&\frac{d\rho_i}{dt}\nonumber\\ 
&& = \frac{\tilde{\lambda}}{2} [ \rho_{i+1} - 2 \rho_i + \rho_{i-1}] 
\nonumber \\
&& - \frac{\tilde \lambda^2}{2} \left[  \theta_2 \frac{\lambda + 2 \tilde \lambda \theta_2}{1-2 \tilde \lambda \theta_2} (\rho_{i+1}^2 - \rho_{i-1}^2)
- (1-2 \theta_2)(\rho_{i+1}^2  - \rho_i^2 ) \right] Fa~~
\nonumber \\
\nonumber \eea 
In the last step,  following 
the assumption of local steady-state, we have used eqs.
(\ref{delta-m1}) and (\ref{delta-m2}) and, subsequently, used the
expression of second moment of local mass as well as the
expression of nearest-neighbor mass-mass correlation
\cite{Das_PRE2016}, \be \langle m_i^2 \rangle =
\frac{1-\tilde{\lambda} (1-2\theta_2)}{\lambda + \tilde{\lambda}
(1-2\theta_2)} \rho_i^2 ~;~ \langle m_{i-1} m_i \rangle = \rho_i^2.
\nonumber \ee

{\it Hydrodynamics.}-- Finally, we take the diffusive limit, by rescaling
space and time as $i \rightarrow x=i/L$, $t \rightarrow \tau =
t/L^2$ and $a \rightarrow 1/L$, and obtain the hydrodynamic
equation for the density field as $\partial_t \rho (x, \tau) +
\partial_x J(\rho(x, \tau)) = 0$ where $J = J_d + J_D$, with
\begin{equation}
J_d(\rho) =  \frac{\tilde{\lambda}^2}{2}\frac{4\theta_2-1}{1-2\tilde{\lambda}\theta_2} \rho^2 ~;~ J_D(\rho) = - \frac{\tilde{\lambda}}{2}\frac{\partial \rho}{\partial x}.
\nonumber
\end{equation}
The above functional forms of currents imply that  the diffusion
coefficient and the conductivity, respectively, have the following
expressions, \bea \chi(\rho) =
\frac{\tilde{\lambda}^2}{2}\frac{4\theta_2-1}{1-2 \tilde \lambda \theta_2}
\rho^2 ~;~ D(\rho) = \frac{\tilde{\lambda}}{2}. \eea The ER as in eq. 
(\ref{ER}) can now be verified by using the previously obtained 
expression of scaled variance \cite{Das_PRE2016},
$$
\sigma^2(\rho) =  \frac{\tilde \lambda (4 \theta_2-1)}{1-2 \tilde \lambda \theta_2} \rho^2.
$$

\section{Density large deviations}
\label{Sec_LDF}

\noindent

The evolution in eq.~ (\ref{continuity}), in fact,  describes the
evolution of the average density profile. As mentioned earlier,
our microscopic models are, however, stochastic by nature, which
gives rise to fluctuations in the density and the associated
current fields. According to the macroscopic fluctuation theory
(MFT) \cite{Bertini_review}, the fluctuations in these two fields
can be introduced by adding a random current field $\zeta(x,\tau)$
to the deterministic one $J(x,\tau)$ as following. The total
current can now be written as $j(x,\tau) = J(x,\tau) + \zeta
(x,t),$ where $\zeta(x,\tau)$ is a weak Gaussian multiplicative
white noise, whose mean is zero and strength
depends on local density through conductivity $\chi(\rho)$,
$$ \langle \zeta(x, \tau) \rangle=0 ~;~\langle \zeta (x,
\tau) \zeta (x', \tau') \rangle = \frac{1}{L} \chi(\rho) \delta
(x-x') \delta (\tau-\tau').$$ Thus, one
obtains the following fluctuating-hydrodynamic time-evolution of
the density field, \bea
\partial_\tau  \rho (x, \tau) + \partial_x \left[ -D(\rho) \partial_x \rho(x,\tau) + \chi (\rho) F + \zeta(x,\tau) \right]=0.~~
\label{FHD} \eea Starting from the stochastic microscopic
dynamics, and using the Markov properties of the evolution, one
can actually prove the above stochastic hydrodynamic equation
(\ref{FHD}) \cite{Bertini_review}. Then, using eq. (\ref{FHD}), one can, in principle,
find the joint probability of any given time-trajectories of the
full density $\rho(x, \tau)$ and current $j(x,\tau)$ profiles,
starting from an arbitrary initial condition.

However, here,  we are interested in the steady-state
probabilities of density large deviations. According to MFT, the
probability of an arbitrary density profile $\rho(x)$ in the
steady state, which corresponds to eq. (\ref{FHD}) with zero
external bias $F=0$, is given by the following large deviation
probability ${\cal P}[\rho(x)] \approx e^{-{\cal F}[\rho(x)]}$,
where the large deviation function ${\cal F}[\rho(x)]$ satisfies
\cite{Bertini_review} \be \int dx~\left[ \partial_x
\left(\frac{\delta \mathcal{F}}{\delta \rho} \right)
\chi(\rho)~\partial_x \left(\frac{\delta \mathcal{F}}{\delta \rho}
\right) - \frac{\delta \mathcal{F}}{\delta \rho} \partial_x
J_D(\rho) \right] = 0.\label{quasipotential} \ee After performing a partial integration
in the second term, one can readily check that the above equation
is satisfied by the LDF $\mathcal{F}[\rho(x)]$ which satisfies the following conditions, \bea
\partial_x\left(\frac{\delta \mathcal{F}}{\delta \rho(x)}\right) &=& \partial_x\left(f'(\rho(x)) - f'(\rho_0)\right),
\label{func-d-LDF}
\\
\frac{1}{f''(\rho) } &=& \frac{\chi(\rho)}{D(\rho)}.
\label{ER-Hyd} \eea Here, $\rho_0$ is  the average or typical
local mass density (which in our case turns out to be the same as
the global density since the systems are homogeneous) at which the
LDF  $\mathcal{F}[\rho]$ has a minimum equal to
$\mathcal{F}[\rho(x)=\rho_0]=0$.  Equation (\ref{func-d-LDF}), together
with this minimum condition, gives the following expression of the
LDF, \be \mathcal{F}[\rho(x)] = \int_{-\infty}^{\infty} dx
\{f(\rho)-f(\rho_0)-f'(\rho_0)(\rho-\rho_0) \}. \label{LDF-h} \ee
Note that the above functional form of the LDF implies the FR as in 
eq. (\ref{f-sigma-rela}). Now substituting eq.
(\ref{f-sigma-rela}) in eq. (\ref{ER-Hyd}), one immediately
obtains the Einstein relation eq. (\ref{ER}). Moreover, using eqs. 
(\ref{ER-Hyd}) and (\ref{f-sigma-rela}), one can easily see that 
the LDF in eq. (\ref{LDF-h}) is exactly the same as in eq. 
(\ref{LDF-a}), which was earlier obtained directly from additivity
and the FR eq. (\ref{f-sigma-rela}). Particularly, for the 
conserved-mass transport processes considered here, one 
recovers free energy density 
$f(\rho)$ and chemical potential $\mu(\rho) = f'(\rho)$, as in Eqs.
\ref{f-rho} and \ref{mu-rho}, by explicitly using the expressions of 
conductivity $\chi(\rho)$ and diffusion coefficients $D(\rho)$ derived in Secs. \ref{Model-I}, \ref{Model-II} and \ref{Model-III}.

\section{results: asymmetric mass transfers}
\label{AMCM}

In the asymmetric mass transport processes, masses are transferred preferentially in a particular direction, say, counter-clockwise. Consequently, there is, on average, a nonzero mass current and detailed balance is manifestly broken in the system. However, even in the case of such asymmetric mass transfer, we explicitly show below that the bulk-diffusion coefficient $D(\rho)$ and the conductivity $\chi(\rho)$ still satisfy an ER. The conductivity (differential) $\chi(\rho) = [\partial J_d/\partial F]_{F=0}$ here can be defined with respect to a small perturbing biasing force field $\vec{F}$ around the nonzero current-carrying steady state. 
For simplicity, only the random sequential update rule is considered here; the results can be straightforwardly generalised to the parallel update rules.

To illustrate how one could incorporate asymmetry in transfer of masses, let us now consider a particular model, say, model I where the dynamics is described by eq. (\ref{model-1}) in Sec. \ref{Model-I}. In this case, model I becomes one having asymmetric transfer of masses, provided that the probability density function $\phi(r_i)$ is not symmetric around $r_i=1/2$. Clearly, the asymmetric mass-transfer gives rise to an inherent bias towards a particular direction. Note that asymmetry can be incorporated in several other ways also, but, for simplicity, we confine our discussions to the cases considered below.

Now, in the above mentioned asymmetric version of model I, the time-evolution of the first moment $\langle m_i (t) \rangle = \rho_i(t)$ of mass at site $i$ is governed by
\begin{eqnarray}
\frac{d\rho_i}{dt}&=& \tilde{\lambda}\langle r_{i-1}m_{i-1} +(1-r_{i+1})m_{i+1} -m_i\rangle \nonumber\\
&&+ \frac{\tilde{\lambda}^2}{2}(2\theta_2 -\theta_1)a\left[\langle m_{i-1}^2\rangle - \langle m_{i+1}^2\rangle\right]F \nonumber\\
&& + \frac{\tilde{\lambda}^2}{2}(2\theta_1 -1)a \left[\langle m_{i+1}^2\rangle - \langle m_{i}^2\rangle\right]F \nonumber\\
\label{asymmetric1}
\end{eqnarray}
Let us define strength of asymmetry $\alpha = [1-2\theta_1]$, which in a particular case may depend on system size $L$ through the first moment $\theta_1$ of probability density function $\phi(r_i)$. The parameter $\alpha$ helps us in obtaining concisely the hydrodynamic equation, which can be applicable to both weakly and strongly asymmetric cases, depending on $\alpha$. 
We now rescale eq. (\ref{asymmetric1}) by $i \rightarrow x=i/L$, $t\rightarrow \tau=t \alpha/L$ and $a\rightarrow 1/L$ and, using the expression $\langle m_{i}^2\rangle = \rho^2_i /[\lambda+2\tilde{\lambda}(\theta_1 - \theta_2)]$ \cite{Das_PRE2016}, we obtain the hydrodynamic equation, 
\be
\frac{\partial \rho}{\partial \tau} = - \tilde{\lambda} \frac{\partial \rho}{\partial x}  +  \nu D \frac{\partial^2 \rho}{\partial x^2} - \frac{\partial}{\partial x}[\nu \chi(\rho) F],
\label{FHD-nu}
\ee
where $\nu=1/\alpha L$ and
$$
\chi(\rho) = \frac{\tilde{\lambda}^2}{2}  \frac{1-4(\theta_1-\theta_2)}{\lambda+2\tilde{\lambda}(\theta_1 - \theta_2)} \rho^2 ~;~~ D(\rho)=\frac{\tilde \lambda}{2}. 
$$
There is now an additional drift current $\tilde{\lambda}\rho$ appearing in the hydrodynamic equation. However, one can immediately verify that, the diffusivity and mobility are indeed connected by the Einstein relation as in eq. (\ref{ER}). 
Note that conductivity now depends on the strength of asymmetry $\alpha$ through $\theta_1=(1-\alpha)/2$.

In the case of weak asymmetry where $\alpha(L) ={\rm const.}/L$ is ${\cal O}(1/L)$, the above rescaling of time ($\tau \sim t/L^2$) leads to diffusive hydrodynamics with conductivity $\nu \chi(\rho)$ and diffusion coefficient $\nu D(\rho)$ both being finite. Whereas, in the case of strong asymmetry where $\alpha = {\rm const.}$ is $ {\cal O}(1)$, the above rescaling of time ($\tau \sim t/L$) gives hyperbolic hydrodynamics with conductivity $\nu \chi(\rho)$ and diffusion coefficient $\nu D(\rho)$ both being infinitesimally small as $\nu \rightarrow 0$ in the hydrodynamic limit. However, the MFT is still expected to describe the density fluctuation in both cases \cite{Bertini_review} and density field $\rho(x,\tau)$ would then satisfy the following stochastic hydrodynamic equation with a Gaussian multiplicative noise-current $\zeta(x,\tau)$,
\be
\frac{\partial \rho}{\partial \tau} = - \partial_x \left[ \tilde{\lambda} \rho  -  \nu D \frac{\partial \rho}{\partial x} + \zeta(x, \tau) \right],
\label{FHD-asym}
\ee
where $ \langle \zeta(x, \tau) \rangle=0$ and $\langle \zeta (x,
\tau) \zeta (x', \tau') \rangle = [\nu \chi(\rho)/L] \delta
(x-x') \delta (\tau-\tau').$ Note that the structure of stochastic hydrodynamics for asymmetric cases remains quite similar to eq. (\ref{FHD}), where $J_D$ is now replaced by $J_D+\tilde \lambda \rho$ and $D$ and $\chi$ are now replaced by $\nu D$ and $\nu \chi$, respectively. Consequently, the density large deviation function can be obtained by solving a slightly modified version of eq. (\ref{quasipotential}),
\bea
\int dx~\left[ \partial_x
\left(\frac{\delta \mathcal{F}}{\delta \rho} \right)
\nu \chi (\rho)~\partial_x \left(\frac{\delta \mathcal{F}}{\delta \rho}
\right) + \frac{\delta \mathcal{F}}{\delta \rho} \partial_x
\nu D (\rho)\partial_x \rho \right]\nonumber
\\
+ \tilde \lambda \int dx \frac{\delta \mathcal{F}}{\delta \rho} \partial_x \rho = 0.~~~
\label{quasipot_asym} 
\eea
By noting that the last term in the l.h.s. of the above equation is identically zero when integration is performed over a periodic boundary, eq. (\ref{LDF-h}), along with eq. (\ref{ER-Hyd}), provides the density LDF, having the same functional form as in eqs. (\ref{LDF-a}), (\ref{f-rho}) and (\ref{mu-rho}). 
One could check that the same LDF can also be recovered by directly using additivity.
The only difference in the two cases of symmetric and asymmetric mass transfers is that  the exact expressions of free energy $f(\rho)$ may differ as it is directly obtained from the ratio (related to parameter $\eta$) of conductivity $\chi(\rho)$ and diffusion coefficient $D(\rho)$ (or, from the mass fluctuation $\sigma^2(\rho)$) and the ratios may be different in these two cases. 
Indeed, the LDFs in the cases of symmetric and strongly asymmetric mass transfer are different as the conductivity $\chi(\rho)$ is different in these two cases. However, the LDFs are the same in symmetric and weakly asymmetric cases, which is somewhat expected.

\begin{figure}[h]
\begin{center}
\includegraphics[width=8.5cm,angle=0.0]{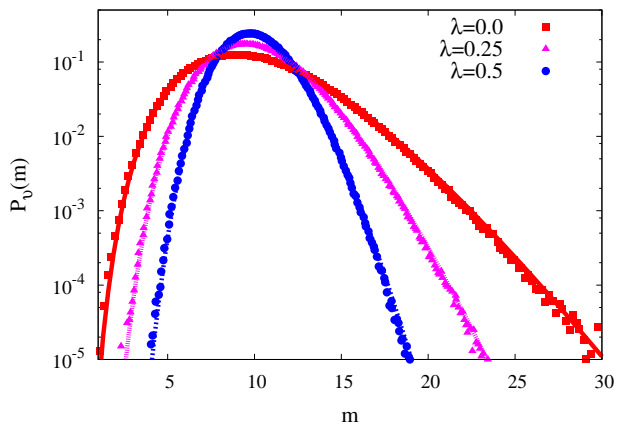}
\end{center}
\caption{Weakly asymmetric mass transfers; Model I (random sequential update): Steady-state probability distribution $P_v(m)$ is plotted as a function of subsystem mass $m$ for $\lambda = 0$, $0.25$ and $0.5$ and subsystem volume $v=10$.}
\label{weak-asym}
\end{figure}

\begin{figure}[h]
\begin{center}
\includegraphics[width=8.5cm,angle=0.0]{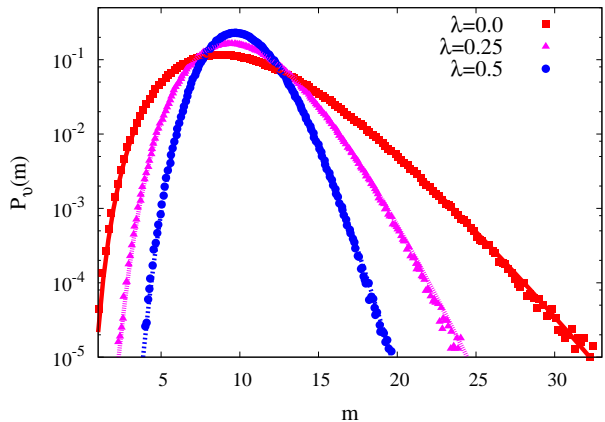}
\end{center}
\caption{Strongly asymmetric mass transfers; Model I (random sequential update): Steady-state probability distribution $P_v(m)$ is plotted as a function of subsystem mass $m$ for $\lambda = 0$, $0.25$ and $0.5$ and subsystem volume $v=10$.}
\label{strong-asym}
\end{figure}

In Figs. and \ref{weak-asym} and \ref{strong-asym}, we have plotted steady-state probability distribution $P_v(m)$  [see eq.(\ref{Pvm})] of mass $m$ in a subsystem of volume $v=10$ as a function of $m$ for $\lambda = 0$, $0.25$ and $0.5$ and $L=5000$, which are in excellent agreement with fluctuating hydrodynamics eq. (\ref{FHD-asym}) as well as additivity property in eq. (\ref{additivity}). It should be noted that, for a particular value of $\lambda$, the subsystem mass distributions are different for weak and strong asymmetry, depending on the parameters $\theta_1$ (or $\alpha$, the strength of asymmetry) and $\theta_2$.

We have also considered asymmetric versions of models II and III, leading to similar conclusions as above (results not presented).

\section{Summary and concluding remarks}
\label{Summary}

In this paper, we have derived hydrodynamics  of paradigmatic
conserved-mass transport processes on a one dimensional ring-geometry,
which have been intensively studied in the last couple of decades.
In these processes, we have calculated two transport coefficients $-$ diffusion coefficient $D(\rho)$ and conductivity $\chi(\rho)$. Remarkably, the two transport coefficients satisfy
an equilibriumlike Einstein relation eq. (\ref{ER}) even when the microscopic dynamics violate detailed balance. In all cases 
studied here, we find that the diffusion coefficient $D$ is
independent of mass density $\rho$ and the conductivity
$\chi(\rho) \propto \rho^2$ is proportional to the square of the mass density $\rho$.
Moreover, using these two transport coefficients, a fluctuating 
hydrodynamic framework for these processes have been set up here, 
following a recently developed macroscopic fluctuation theory (MFT). 
The MFT has helped us to calculate density large deviation function 
(LDF), which is analogous to an equilibriumlike free energy density 
function. The LDFs completely agree with that obtained previously in Refs. \cite{Chatterjee_PRL2014, Das_PRE2016} solely using an additivity
property eq. (\ref{additivity}).

Interestingly, the analytically obtained functional dependence  of the two
transport coefficients $D(\rho)$ and $\chi(\rho)$ on density
indicates that, on large space and time scales, these mass
transport processes belong to the class of
Kipnis-Marchioro-Presutti (KMP) processes. However, unlike the KMP
processes on a ring, the processes studied in this paper
generally have a nontrivial spatial structure in their steady
states. That is, they have finite spatial correlations in the
steady state. Not surprisingly, the exact probability weights of microscopic configurations in the steady-state, except for a few special cases \cite{Majumdar_PRE1996, Krug_JSP2000, Rajesh_JSP2000, Zielen_JSP2002}, are not yet known.
In fact, precisely due to this nontrivial spatial steady-state structure in out-of-equilibrium interacting-particle systems, finding hydrodynamics in such systems poses a great challenge. 
This is because, in the absence of
knowledge of the exact steady-state weights, it is usually
difficult to calculate averages of local observables (e.g.,
moments of local mass variables, which have been actually used
here to derive hydrodynamics of these processes).

However, as noted in Ref. \cite{Das_PRE2016}, there is an
important feature in these conserved-mass transport processes
(with zero external bias $F=0$), arising from the fact that the
Bogoliubov-Born-Green-Kirkwood-Yvon (BBGKY) hierarchy involving
$n-$point spatial correlations in the steady states closes. In
other words, $n-$point spatial correlations in the steady state do
not depend on $(n+1)-$point or any higher-order spatial
correlations. This particular property previously enabled us to
exactly calculate the steady-state $2-$point spatial correlations
and, consequently, the second moment $\langle m_i^2 \rangle$ of
local mass at site $i$ \cite{Das_PRE2016}. Indeed, the  second
moment of local mass, which appears in the hydrodynamic equations
[e.g., see eq. (\ref{density_model1})], determines the functional
dependence of the conductivity $\chi(\rho)$ on density $\rho$.

Finally, it is worth mentioning that, unlike in equilibriumm, microscopic dynamics in the mass transport processes considered here in general do not satisfy detailed balance. Even for the processes with symmetric mass transfers, we have explicitly shown that Kolmogorov criterion, and thus detailed balance, is violated (even in the absence of a biasing force) and the microscopic dynamics is not time reversible. That is, for a forward path in the configuration space, there may not exist a reverse path. 
However, in spite of the lack of any microscopic reversibility in the dynamics of the processes, the observed Einstein relation suggests that
these mass transport processes possess a kind of time-reversibility
on a coarse-grained macroscopic (hydrodynamic) level. As discussed here, this
macroscopic time-reversibility can be understood in the light of
a macroscopic fluctuation theory (MFT) \cite{Bertini_review}, which indeed correctly predicts the probabilities of density large-deviations obtained earlier in Refs.
\cite{Chatterjee_PRL2014, Das_PRE2016}. From an overall
perspective, we believe our study could provide some useful 
insights in characterizing fluctuations in many other driven many-
particle systems, e.g., various driven lattice gases
\cite{Chatterjee_PRL2014, Book_Spohn}, where a
fluctuating hydrodynamic description is yet to be obtained.

\section{Acknowledgement}

PP gratefully acknowledges the Science and Engineering Research Board (SERB), India, under Grant No. EMR/2014/000719, for financial and computational support throughout the work. 
AK acknowledges the SERB, India (under the above Grant) as well as the S. N. Bose National Centre for Basic Sciences, Kolkata for the hospitality provided during a visit where a part of the work was done.
This work was initiated at the International Centre for 
Theoretical Sciences (ICTS), Bengaluru where AK and PP both participated in a workshop on ``Non-equilibrium statistical physics'' 
(Code: ICTS/Prog-NESP/2015/10).

\end{document}